\documentclass[manuscript=article,layout=twocolumn]{achemso}
\usepackage[version=3]{mhchem} 
\usepackage[]{hyperref}
\usepackage[nameinlink,noabbrev]{cleveref}
\usepackage{siunitx}
\usepackage{xcolor}
\usepackage{sansmath} 
\sansmath 
\newcommand{\ma}[0]{\ce{MAPbI_{3}}}
\newcommand{\supercell}[3]{$#1\times#2\times#3$}

\author{Viren Tyagi}
    \affiliation{Department of Applied Physics and Science Education, Eindhoven University of Technology, 5600 MB, Eindhoven, The Netherlands}
\author{Geert Brocks}
    \affiliation{Department of Applied Physics and Science Education, Eindhoven University of Technology, 5600 MB, Eindhoven, The Netherlands}
    \alsoaffiliation{Computational Chemical Physics, Faculty of Science and Technology and MESA+ Institute for Nanotechnology, University of Twente, 7500 AE, Enschede, The Netherlands}
\author{Shuxia Tao}
    \affiliation{Department of Applied Physics and Science Education, Eindhoven University of Technology, 5600 MB, Eindhoven, The Netherlands}
    \email{s.x.tao@tue.nl}

\title{A unified microscopic picture of cation and anion migration in \ce{MAPbI_{3}}}


\begin{document}

\begin{abstract}

Passivating defects and restricting defect mobilities in halide perovskites to increase device lifetimes has become a main field of research. Modeling structure and mobility of point defects is an essential contribution to this endeavor. We employ molecular dynamics, based on neural network potentials trained on density functional theory data, to model ion migration in MAPbI$_3$ triggered by \ce{I} and \ce{MA} vacancies or interstitials. Most of these species diffuse rapidly at room temperature, with migration barriers between 0.15 and 0.20 eV. \ce{MA} interstitials are highly mobile despite their molecular nature, owing to a concerted migration mechanism involving multiple \ce{MA} ions. No evidence of \ce{MA} vacancy migration is obtained. Whereas diffusion of \ce{I}-related defects appreciably depends on their charge state, diffusion of \ce{MA} defects does not. These results revise the conventional picture of ion transport in hybrid perovskites and highlight the role of collective molecular motion in enabling fast ionic migration.

\end{abstract}

\section{Introduction}
Hybrid organic-inorganic metal halide perovskites have shown great potential for application in photovoltaic and light-emitting devices \cite{Shen2026key}. The materials allow for optimization of optoelectronic properties through chemical compositional flexibility and low temperature solution processing \cite{Yi2016entropic,Bag2016large}. Halide perovskites are soft materials, however, where defects in the crystal lattice are relatively easily formed \cite{Ball2016defects,Wang2018defects}. Whereas most of these defects do not seem to alter the basic optical and electronic parameters much, they can trigger chemical decomposition and degradation \cite{Thiesbrummel2024ion}. Defects facilitate ion migration through the material to surfaces, grain boundaries, and interfaces with electrodes, where degradation reactions are likely to take place \cite{apergi2022decomposition,apergi2024probing}. Passivating defects or limiting defect mobilities in order to enlarge device lifetimes has become a main field of research in halide perovskites \cite{Semerci2025role,Maschwitz2025crystallization}. Identifying the character and mobility of the relevant migrating defects is then the first essential step. 

A majority of experimental studies on ion transport in halide perovskites focus on MAPbI$_3$ (MA = \ce{CH_{3}NH_{3}^{+}}, methylammonium) as the prototype halide perovskite material. Assuming that the observed ion migration behaviors relate to intrinsic properties of bulk MAPbI$_3$, the prime vehicles for diffusion should be point defects, i.e., single ion interstitials and/or vacancies, rather than extended or compound defects \cite{Xue2023compound}. It is generally agreed upon that \ce{Pb} interstitials and vacancies are immobile at room temperature, which leaves \ce{I} and/or \ce{MA} defects as possibilities \cite{clark2020formation,Thiesbrummel2026ion}.

 Halide anion migration receives most attention in MAPbI$_3$. Temperature-dependent measurements allow for the extraction of migration barriers, but the numbers extracted from different experiments show a considerable spread. Experiments cover (temperature-dependent) conductivity, capacitance or photoluminescence, \cite{hoke_reversible_2014,eames_ionic_2015,yang_significance_2015,yuan_photovoltaic_2015,yu_native_2016,li_iodine_2016,mosconi_light-induced_2016,dequilettes_photo-induced_2016, xing_ultrafast_2016, game_ions_2017,awni_influence_2020,schmidt_quantification_2025} transient ion drift,\cite{futscher_quantification_2019} impedance and deep-level transient spectroscopy, \cite{bag_kinetics_2015,pockett_microseconds_2017,futscher_quantifying_2020,reichert_probing_2020,mcgovern_grain_2021} and find diffusion barriers ranging from 0.10 to 0.68 eV. Some of these experiments also report barriers for the diffusion of MA cations, with again a considerable spread in numbers from 0.18 to 0.94 eV. Per study the numbers for the migration barriers of MA cations tend to be somewhat larger than those for the I anions, but both species can be mobile. 

Whereas charge states, energy levels, and diffusion coefficients of migrating species are experimentally accessible, the nature and composition of the actual migrating defects are less so. Studies therefore use density functional theory (DFT) calculations as support. The workhorse method for extracting migration barriers is based on transition state theory (TST) and finding a lowest-energy path following the nudged elastic band (NEB) method, for instance. This approach suffers from the difficulty of finding such a path in a soft material such as MAPbI$_3$, where in fact many such paths may compete and contribute, as the low symmetry of the MA cation enables multiple energetically competitive configurations \cite{bokdam2017assesing,lahnsteiner2018finite}. Values reported for the migration barriers range from 0.08 to 0.58 eV for the iodide vacancy, 0.08 eV for the iodide interstitial, and 0.46 to 0.96 eV for the MA vacancy.\cite{eames_ionic_2015,haruyama2015first,azpiroz2015defect,Pazoki2017vacancy,Ferdani2019partial,Woo2022factors,zhao2024fast} We do not know of reports on such a calculation on the MA interstitial, although its formation energy is predicted to be only moderate to low.\cite{xue2021first,xue2022intrinsic}  

Migration rates or diffusion coefficients cannot be extracted directly from a NEB calculation. This is possible using molecular dynamics (MD) simulations. Molecular dynamics samples all relevant migration pathways and configurations. It allows for direct simulations of ion diffusion at preset temperatures, giving access to diffusion rates and barriers. The Achilles heel of MD is the force field used in the simulations. From a parametrized analytical force field, migration barriers of 0.1 and 0.55 eV have been extracted for the iodide vacancy and interstitial, respectively.\cite{delugas2016thermally,Balestra2020efficient} The accuracy of such force fields is debated, however.\cite{lahnsteiner2018finite} A breakthrough has been achieved in recent years in the form of machine learned force fields (MLFFs), which allow for MD simulations using force fields at DFT accuracy \cite{Behler2021four,unke2021machine,wu2023applications}. By selecting the DFT data on which the MLFFs are trained, it is possible to target MD simulations toward specific charge states of the migrating species \cite{tyagi_tracing_2025,mosquera-lois2025point}.   

We use MD based on MLFFs to model ion migration in MAPbI$_3$ triggered by I anion vacancies or interstitials, and by MA cation vacancies or interstitials. We find all species to be extremely mobile, except for the MA vacancy. The mobility of halide anion defects, vacancies and interstitials, is comparable to that found in the all inorganic compounds CsPb(I$_x$Br$_{1-x}$)$_3;\; x=0,...,1$ \cite{tyagi_tracing_2025,tyagi2026halide}, leading to the conclusion that the A cation species in the perovskite structure does not affect halide motion much. Remarkably, the diffusion rate of MA interstitial cations is only slightly smaller than that of I interstitial anions.  Migration of MA interstitals proceeds via a concerted motion involving the interstitials and two of its MA neighbors in the lattice. Whereas the diffusion of I-related defects very much depends on their charge state, that of MA-related defects is virtually independent of charge state. We attribute this to the fact that an MA cation interstitial or vacancy only create very shallow impurity levels that are ionized at room temperature. 

\section{Methods}
Iodide and MA interstitials and vacancies have charge states $\mathrm{I_{I}^{-}}$, $\mathrm{V_{I}^{+}}$, $\mathrm{I_{MA}^{+}}$, and $\mathrm{V_{MA}^{-}}$, respectively, under intrinsic or mildly doped conditions \cite{meggiolaro2018first,xue2022intrinsic}. Results obtained from DFT calculations differ somewhat on whether vacancies or interstitials are predicted to be the dominant species under equilibrium conditions. This reflects the use of different functionals, for instance \cite{meggiolaro2018first,xue2021first}. 
The formation energy of interstitial-vacancy (Frenkel) pairs $\mathrm{I_{I}^{-}}/\mathrm{V_{I}^{+}}$ and $\mathrm{I_{MA}^{+}}/\mathrm{V_{MA}^{-}}$ only weakly depends on the functional \cite{xue2021first}, However, including Vanderwaals (VdW) interactions favors interstitials over vacancies, whereas excluding VdW interactions favors vacancies, demonstrating that VdW interactions have to be included \cite{meggiolaro2018first,xue2021first}. 

In this paper we consider both vacancies and interstitials and train a MLFF for describing them in their most prominent charge states under intrinsic conditions. Our MLFF is based on the Allegro model, which is a local E(3)-equivariant graph neural network potential (NNP), as implemented in the NequIP package \cite{Batzner2022equivariant,Musaelian2023learning}. The datasets for training and validating the network are generated using the on-the-fly module as implemented in VASP, sampling structures from short-time scale MD runs \cite{Kresse1996efficient,Jinnouchi2019onthefly,Jinnouchi2019phase,bartok2010gaussian,bartok2013onrepresenting}. Energies, forces, and stress tensors are calculated using DFT with the SCAN+rVV10 functional.\cite{Sun2015strongly,Sabatini2013nonlocal} These are calculated for a series of structures that are sufficiently different from one another, using selection criteria based on Bayesian inference. The training and validation sets are generated using \supercell{2}{2}{2} cubic supercells \ma (96 atoms) containing one charged or neutral \ce{MA} or \ce{I} point defect. Details of the training procedure are explained in the SI note 1.

Iodide vacancies (interstitials) give acceptor (donor) levels that can be populated with electrons (holes) if the (quasi) Fermi level is very close to the conduction (valence) band, which might happen under device operating conditions \cite{Motti2019controlling,Ni2022evolution}. The same is true in principle for MA  interstitials and vacancies. A NNP trained on energies and forces has no direct knowledge of electronic structure, but it is possible to select DFT data for training a neural network that are generated with a Fermi level in or close to the valence or conduction band to represent defects with modified charge states. In this way, we train separate NNPs for $\mathrm{I_{MA}^{0}}$ and $\mathrm{V_{MA}^{0}}$.

\begin{figure}
    \includegraphics{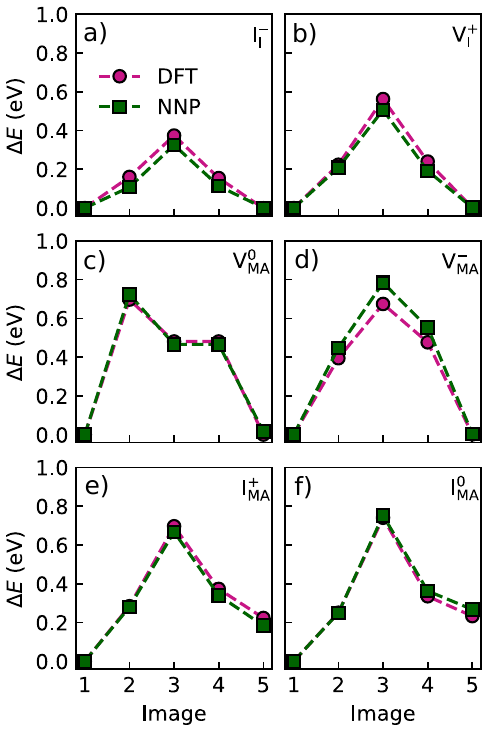}
    \caption{{ Energies along the NEB migration paths, calculated using DFT and the NNPs. The points are the calculated values, and the lines guide the eye; the energy of the minima is set to 0.}}
    \label{fig:NEB_energies}
\end{figure}

The accuracy of the NNPs is explored by comparing the migration barriers along paths generated using the climbing image nudged elastic band (CI-NEB) technique with those calculated by DFT along the same paths \cite{henkelman2000climbing}. The computed energies are given in Figure~\ref{fig:NEB_energies}, and the computed migration barriers in Table~\ref{tab:NEB_energies}. The difference between the migration barriers calculated using DFT and the NNP ($\mathrm{|\Delta E_{b}^{DFT}-\Delta E_{b}^{NNP}|}$) is less than 0.11 eV in all cases, indicating the accuracy of the NNPs. No migration events were actually observed during training for the neutral MA vacancy, $\mathrm{V_{MA}^{0}}$, yet the NNP accurately describes the migration barrier for this defect ($\mathrm{|\Delta E_{b}^{DFT}-\Delta E_{b}^{NNP}|\,=\,}$0.02 eV) (Figure~\ref{fig:NEB_energies}c), demonstrating the accuracy of the NNP in extrapolating to events not explicitly included in training. 

For $\mathrm{I_{MA}^{+}}$ (Figure~\ref{fig:NEB_energies}e) and $\mathrm{I_{MA}^{0}}$ (Figure~\ref{fig:NEB_energies}f), the initial and the final images do not have the same energy, indicating that these are not the complete migration paths. It illustrates the problem of finding the lowest energy migration path in this compound. As MA cations are asymmetric and polar, the local structure around a migrating I defect allows for multiple local minima, and significant hysteresis. The details of the CI-NEB calculations are given in the SI note 2. 

The forces calculated using the NNP are also compared with those calculated using DFT. This is done using structures sampled from MD simulations at 600 K performed on \supercell{6}{6}{6} supercells of \ma (2592 atoms) with one \ce{I} or \ce{MA} point defect. From these comparisons the coefficient of determination $\mathrm{R^{2}=0.99}$ and the mean absolute error (MAE) $\mathrm{\leq 49.06\,meV/\AA}$. Specifically for the forces on atoms in the defect environment we find $\mathrm{R^{2}\,\geq\,0.98}$ and $\mathrm{\leq 64.38\,meV/\AA}$, indicating the high accuracy of the NNP, see SI note 2.

\begin{table}[]
    \centering
    \begin{tabular}{|c c c|}
        \hline
        System & $E_{\mathrm{b}}^{\mathrm{DFT}}$ (eV) & $E_{\mathrm{b}}^{\mathrm{NNP}}$ (eV) \\
        \hline
        $\mathrm{{I_{I}^{-}}}$ & {0.37} & {0.33} \\
        $\mathrm{{V_{I}^{+}}}$ & {0.56} & {0.51} \\
        $\mathrm{{V_{MA}^{0}}}$ & {0.70} & {0.72} \\
        $\mathrm{{V_{MA}^{-}}}$ & {0.67} & {0.78} \\
        $\mathrm{{I_{MA}^{+}}}$ & {0.70} & {0.67} \\
        $\mathrm{{I_{MA}^{0}}}$ & {0.74} & {0.75} \\
        \hline
    \end{tabular}
    \caption{\textmd{Defect migration barriers ($E_\mathrm{{b}}$) calculated using DFT and the NNPs for the NEB migration paths.}}
    \label{tab:NEB_energies}
\end{table}

\section{Results}
Following training and validation, the NNPs are used to perform three independent 2 nanosecond (ns) long MD simulations at each temperature for five temperatures between 500 K and 600 K. The details of these production runs are given in the SI note 4. From these simulations, diffusion coefficients are extracted, providing the temperature-dependent diffusion behavior. As ion transport in MAPb$_3$ via interstitials or vacancies takes place through kick-out processes, tracer diffusion is not appropriate for characterizing transport. Instead we include all atoms or molecules of a particular species in calculating the diffusion coefficient, using the Einstein relation, which links the slope of the mean square displacement to time

\begin{multline}
        D = \frac{1}{6}\lim_{t\rightarrow\infty}\frac{d}{dt}
        \Bigl\langle \sum_{i=1}^{N} 
        \left\Vert \mathbf{r}_i(t+t_0)-\mathbf{r}_i(t_0)\right\Vert^{2}
        \Bigr\rangle_{t_{0}}; \\
         t > t_0
    \label{eqn:equation_MSD}
\end{multline}

where $N$ is the number of atoms of a particular species in the simulation box, and $\mathbf{r}_i$ are the atomic positions. Notwithstanding the fact that all ions participate in extracting the diffusion coefficient $D$, in the following we will still label them according to the defect, interstitial or vacancy, that instigates the diffusion.  

\begin{figure}
    \includegraphics{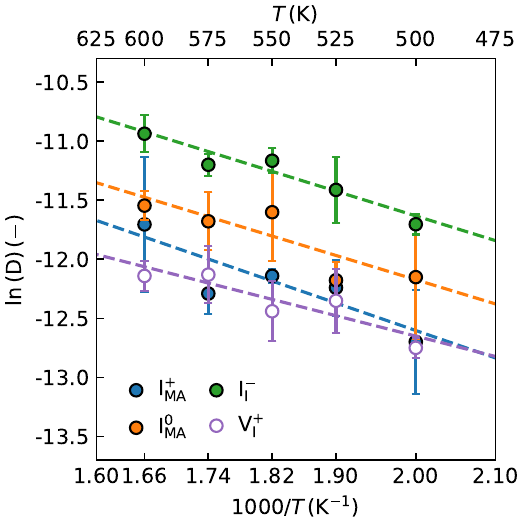}
    \caption{Temperature-dependent diffusion coefficients of iodide interstitial ($\mathrm{I_{I}^{-}}$), iodide vacancy ($\mathrm{V_{I}^{+}}$), neutral \ce{MA} interstitial ($\mathrm{I_{MA}^{0}}$), and positive \ce{MA} interstitial ($\mathrm{I_{MA}^{+}}$) defects in \ma. The dashed lines represent the fits to an Arrhenius expression, and the error bars represent the standard error in mean at each point.}
    \label{fig:diffusion-curve}
\end{figure}

The diffusion coefficients $D$ extracted from the simulations are given in Figure~\ref{fig:diffusion-curve}. For iodide interstitial $\mathrm{I_{I}^{-}}$ and vacancy $\mathrm{V_{I}^{+}}$, the diffusion coefficients are within a factor of five of one another within the given temperature range, with the interstitial being the more mobile of the two. Remarkably, also the diffusion coefficients of the neutral and positive \ce{MA} interstitials $\mathrm{I_{MA}^{0}}$, $\mathrm{I_{MA}^{+}}$ are within the same range. The \ce{MA} vacancies $\mathrm{V_{MA}^{0}}$, $\mathrm{V_{MA}^{-}}$ are found to be immobile in the simulated temperature and time ranges.

For all defects the temperature dependence of the diffusion coefficient can be fitted to an Arrhenius relation

\begin{equation}
    D = D_{0}\exp\left(-\frac{E_\mathrm{a}}{k_\mathrm{B}T}\right)
    \label{eqn:equation1}
\end{equation}

where $k_\mathrm{B}$ is the Boltzmann constant, $E_\mathrm{a}$ the activation energy, and $D_{0}$ the pre-exponential factor. The fits are given in Figure~\ref{fig:diffusion-curve}, and the fit parameters along with the extrapolated diffusion coefficient at room temperature ($D_{\mathrm{300K}}$) are presented in Table~\ref{tab:diffusion-data}. 

\begin{table*}[htbp!]
    \centering
    \begin{tabular}{|c c c c|}
        \hline
         System & ${E_\mathrm{a}\,\,\mathrm{(eV)}}$ & ${D_{0}\,\,(\times10^{-4}\,\mathrm{cm}^{2}\mathrm{s}^{-1})}$ & ${D_{300K}\,\,(\times10^{-7}\,\mathrm{cm}^{2}\mathrm{s}^{-1})}$\\
        \hline
         $\mathrm{I_{I}^{-}}$ & {$\mathrm{0.18\pm0.03}$} & {$\mathrm{5.93\pm3.97}$} & {$\mathrm{5.61}$}\\
         $\mathrm{V_{I}^{+}}$ & {$\mathrm{0.15\pm0.04}$} & {$\mathrm{1.04\pm0.92}$} & {$\mathrm{3.19}$}\\
         $\mathrm{I_{MA}^{0}}$& {$\mathrm{0.18\pm0.06}$} & {$\mathrm{3.16\pm4.27}$} & {$\mathrm{2.99}$}\\
         $\mathrm{I_{MA}^{+}}$ & {$\mathrm{0.20\pm0.07}$} & {$\mathrm{3.47\pm5.12}$} & {$\mathrm{1.51}$}\\ 
         \hline
    \end{tabular}
    \caption{\textmd{Activation energies (${E_\mathrm{a}}$) and pre-exponential factors (${D_{0}}$) extracted from the Arrhenius fits, and extrapolated diffusion constants (${D_{300\mathrm{K}}}$) at room temperature for iodide interstitial ($\mathrm{I_{I}^{-}}$), iodide vacancy ($\mathrm{V_{I}^{+}}$), neutral \ce{MA} interstitial ($\mathrm{I_{MA}^{0}}$), and positive \ce{MA} interstitial ($\mathrm{I_{MA}^{+}}$) in \ma.}}
    \label{tab:diffusion-data}
\end{table*}

The activation energies and prefactors for interstitial $\mathrm{I_{I}^{-}}$ and vacancy, $\mathrm{V_{I}^{+}}$ are quite close, and their extrapolated diffusion coefficients at 300 K are within a factor of two. The numbers are also quite close to those found for halide interstitials in CsPbI$_3$, CsPbBr$_3$ and CsPb(I$_x$Br$_{1-x}$)$_3$ \cite{tyagi_tracing_2025,tyagi2026halide}, indicating that halide diffusion in 3D perovskites does not depend too much on the A cation in the lattice.

Diffusion coefficients for \ce{MA} interstitials have similar prefactors as those for I defects, and only slightly higher activation energies, which demonstrates that \ce{MA} interstitials are as mobile as \ce{I} defects. In addition, the charge state of the \ce{MA} interstitial does not seem to impact its migration behavior much. A neutral \ce{MA} interstitial $\mathrm{I_{MA}^{0}}$ creates one donor level very close to the conduction band edge \cite{meggiolaro2018first,xue2021first}, implying that at room or higher temperatures the MA interstitial will be ionized $\mathrm{I_{MA}^{+}}$. Indeed, a charge density analysis shows that for $\mathrm{I_{MA}^{0}}$ the extra electron delocalizes over the whole lattice (SI note 5), such that it does not impact the motion of the MA interstitial much. 

From the fact that \ce{MA} vacancies are immobile during a 2 ns simulation at the highest temperature considered here (600 K), one can estimate a lower bound for their activation energy $E_\mathrm{a}$. Assuming Arrhenius behavior with a prefactor similar to the other defects $D_0\approx 10^{-4}$ cm$^2$s$^{-1}$, gives an attempt frequency $\nu_0 \approx 0.5$ THz \cite{tyagi_tracing_2025}. The number of migration events in 2 ns then drops below one for $E_\mathrm{a} \approx 0.4$ eV.

To gain atomistic insights into the diffusion behaviour of these defects, trajectories are analyzed as defects migrate. Iodide defects migrate very similar to the way they do in \ce{CsPbI_{3}}\cite{tyagi_tracing_2025}. Vacancies $\mathrm{V_{I}^{+}}$ move through a kick-out mechanism with a neighboring \ce{I} atom rotating $90^\circ$ around the \ce{Pb-I} bond filling in the vacancy. The migration path for $\mathrm{I_{I}^{-}}$ interstitials consists of hopping moves of an \ce{I} atom from one \ce{Pb-I, I-Pb} double bridge to a neighboring \ce{Pb-I-Pb} bond to form a double bridge there. 

\begin{figure*}[htbp!]
    \includegraphics{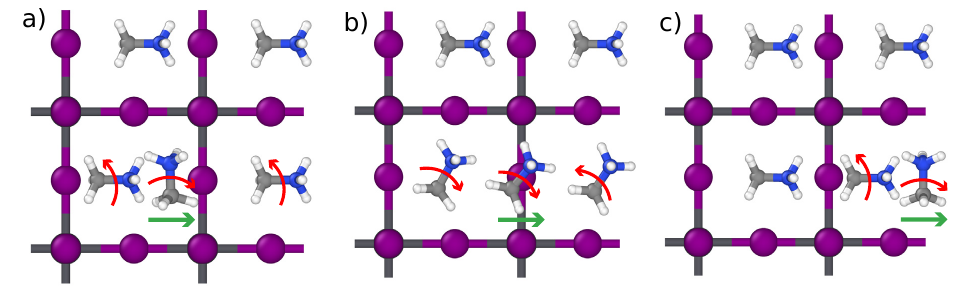}
    \caption{Schematic representation of the diffusion of the \ce{MA} interstitial. The red arrows represent the rotation of the \ce{MA} ion along the \ce{C-N} axis, and the green arrows represent the migration paths.}
    \label{fig:migration_schematic}
\end{figure*}

The \ce{MA} interstitial migrates through the concerted motion of three neighboring \ce{MA} ions (Figure~\ref{fig:migration_schematic}). The \ce{MA} interstitial exists most frequently as two \ce{MA} ions in one \ce{Pb-I} cage with their \ce{C-N} axes oriented approximately perpendicular to each other (Figure~\ref{fig:migration_schematic}a). Migration is instigated by these two \ce{MA} ions and a neighboring \ce{MA} ion rotating such that their \ce{C-N} axes align approximately parallel to each other (Figure~\ref{fig:migration_schematic}b). Finally, the \ce{MA} ion in the center migrates from one \ce{Pb-I} cage to the neighboring one (Figure~\ref{fig:migration_schematic}c).

\section{Discussion}
To place our results in context, we compare the calculated migration barriers (Table~\ref{tab:diffusion-data}) with the wide range of experimentally reported activation energies (Table~\ref{tab:expE_a}) for ion migration in \ce{MAPbI_{3}}. Most experimental data gathered on ion migration in MAPbI$_3$ have been assigned to halide migration\cite{yantara2024toolsets}, with reported activation energies from 0.10 to 0.68 eV, extracted with different measurement techniques, including conductivity/current measurements\cite{yang_significance_2015,li_iodine_2016,xing_ultrafast_2016,game_ions_2017}, capacitance/impedance measurements\cite{pockett_microseconds_2017,futscher_quantifying_2020,reichert_probing_2020,mcgovern_grain_2021,awni_influence_2020}, photoluminescence \cite{hoke_reversible_2014,mosconi_light-induced_2016,dequilettes_photo-induced_2016}, photocurrent relaxation\cite{eames_ionic_2015}, current response measurements\cite{yu_native_2016}, ionic current measurements\cite{schmidt_quantification_2025}, and transient ion drift \cite{futscher_quantification_2019}. Some experiments also report results assigned to the diffusion of MA cations with barriers between 0.23 and 0.94 eV, from conductivity/current measurements\cite{yuan_photovoltaic_2015,game_ions_2017}, impedance measurements\cite{bag_kinetics_2015,pockett_microseconds_2017}, deep-level transient spectroscopy\cite{reichert_probing_2020}, and ionic current measurements\cite{schmidt_quantification_2025}. Apart from technical issues and interpretational difficulties regarding the experimental data,\cite{yantara2024toolsets,schmidt2024consistent} it is clear that there is a considerable spread in these numbers. 

\begin{table*}[]
    \centering
    \begin{tabular}{|c c c|}
        \hline
        Technique & Activation energy (eV) & Migrating Species \\
        \hline
        Bulk conductivity\cite{yang_significance_2015} & 0.43 & - \\
        Conductivity\cite{yuan_photovoltaic_2015} & 0.36 & $\mathrm{MA^{+}}$ ion \\
        Current density\cite{li_iodine_2016} & 0.23-0.31 & $\mathrm{I_{I}^{-}}$ \\
        Conductivity\cite{xing_ultrafast_2016} & 0.08-0.14 and 0.47 & - \\
        Current density\cite{game_ions_2017} & 0.10 and 0.41 & $\mathrm{V_{I}^{+}}$ and $\mathrm{V_{MA}^{-}}$ \\
        Ionic current\cite{schmidt_quantification_2025} & 0.28 and 0.94 & $\mathrm{V_{I}^{+}}$ and $\mathrm{V_{MA}^{-}}$ \\
        Current response\cite{yu_native_2016} & 0.10-0.18 & $\mathrm{V_{I}^{+}}$ \\
        \hline
        Impedance spectroscopy\cite{bag_kinetics_2015} & 0.58 & $\mathrm{MA^{+}}$ ion \\
        Impedance spectroscopy\cite{pockett_microseconds_2017} & 0.55 and 0.68 & $\mathrm{V_{I}^{+}}$ and $\mathrm{V_{MA}^{-}}$ \\
        Impedance spectroscopy\cite{futscher_quantifying_2020} & 0.13 & halide ions \\
        Impedance spectroscopy\cite{reichert_probing_2020} & 0.18 & $\mathrm{I_{MA}^{+}}$ \\
        Impedance spectroscopy\cite{mcgovern_grain_2021} & 0.50 & halide anions \\
        Capacitive signature\cite{awni_influence_2020} & 0.36 & - \\
        Deep-level transient spectroscopy\cite{reichert_probing_2020} & 0.23 and 0.14 & $\mathrm{I_{MA}^{+}}$ and $\mathrm{I_{I}^{-}}$ \\
        \hline
        Photoluminescence rise rates\cite{hoke_reversible_2014} & 0.27 & halide ions \\
        Photoluminescence rise rates\cite{mosconi_light-induced_2016} & 0.14 & Iodide defects \\
        Photoluminescence rise rates\cite{dequilettes_photo-induced_2016} & 0.19 & Iodide defects \\
        \hline
        Photocurrent relaxation\cite{eames_ionic_2015} & 0.60-0.68 & $\mathrm{V_{I}^{+}}$ \\
        \hline
        Transient ion-drift\cite{futscher_quantification_2019} & 0.29 and 0.39-0.90 & $\mathrm{I^{-}}$ ion and $\mathrm{MA^{+}}$ ion \\
        \hline
    \end{tabular}
    \caption{\textmd{Experimentally reported activation energies for defect migration.}}
    \label{tab:expE_a}
\end{table*}

Calculated activation barriers are typically based on NEB calculations, with numbers
reported for the iodide vacancy of 0.08-0.58 eV, the iodide interstitial 0.08 eV, the MA vacancy 0.46-0.96 eV.\cite{eames_ionic_2015,haruyama2015first,azpiroz2015defect,Pazoki2017vacancy,Ferdani2019partial,Woo2022factors,zhao2024fast} The considerable spread in those numbers is associated with the difficulties of finding the lowest energy path for diffusion in a soft material such as MAPbI$_3$, where moreover the asymmetry of the MA cation may give rise to multiple competing paths. 

Our MD simulations sample all these paths according to the appropriate thermal distribution, and yield a consistent set of activation energies in a narrow range of 0.15-0.20 eV, thereby supporting the lower values of the experimental spread. One should remark that in order to gather sufficient statistics regarding migration events, we have performed the simulations at elevated temperatures, $T\geq500$ K, and assume that the Arrhenius expression for the diffusion coefficient, Eq. \ref{eqn:equation1}, holds for temperatures down to room temperature. This assumption might in principle be tested using rare-event sampling techniques, but that would involve a considerable computational effort outside the scope of the present paper. 

As for the identification of the moving halide defect, from the simulations there is little difference in the mobility of the iodide vacancy or interstitial. Under equilibrium conditions the interstitial is the most favored point defect,\cite{meggiolaro2018first,xue2021first} but under non-equilibrium conditions the vacancy might of course occur. 

As for the MA defect, there is a clear distinction between interstitial and vacancy, where the simulation predicts the MA vacancy to be immobile within the simulated temperature and time scales, whereas the MA interstitial has a mobility comparable to that of the iodide defects. The latter challenges the common assumption that \ce{MA} migration is intrinsically slow due to its molecular nature. Our calculated migration barrier of 0.20 eV for $\mathrm{I_{MA}^{+}}$ (Table~\ref{tab:diffusion-data}) is in agreement with the values obtained experimentally by Reichert et al. (Table~\ref{tab:expE_a}) using impedance spectroscopy (0.18 eV) and deep-level transient spectroscopy (0.23 eV)\cite{reichert_probing_2020}. Again, equilibrium conditions favor the formation of interstitials,\cite{meggiolaro2018first,xue2021first} with the MA vacancy only occurring under very iodine-rich conditions. 

Some of the spread in experimental numbers might be explained by a difference in time and length scales caused by the inhomogeneity of the samples. These are typically multi-crystalline thin films, where diffusion in a crystalline grain will be different from that at grain boundaries/surfaces or between grains. As our simulations only consider bulk diffusion, this might suggest that the higher numbers observed in experiments refer to inter-grain diffusion.

\section{Conclusions}
In summary, we study the diffusion of iodide and \ce{MA} interstitials and vacancies in \ce{MAPbI_{3}}. We trained a neural network potential for the most stable charge states of these defects ($\mathrm{I_{I}^{-}}$,$\mathrm{I_{MA}^{+}}$,$\mathrm{V_{I}^{+}}$, and $\mathrm{V_{MA}^{-}}$), and separate neural network potentials for neutral \ce{MA} interstitial ($\mathrm{I_{MA}^{0}}$) and vacancy ($\mathrm{V_{MA}^{0}}$) using DFT calculations. Using these force fields, we performed MD simulations to study the migration behaviour of these defects.

We find that all defects besides \ce{MA} vacancies are mobile with migration barriers in a narrow range of 0.15-0.20 eV, consistent with the lower end of experimentally reported activation energies. The migration rates of the \ce{MA} interstitials are very similar to those of iodide point defects, despite \ce{MA} interstitials being much bulkier than iodide defects. We attribute this finding to the concerted motion of three neighboring MA ions facilitating \ce{MA} interstitial migration. In addition, our findings suggest that unlike iodide point defects, where the migration rates depend on the defect charge state, those of \ce{MA} defects are not impacted by the charge state.

These findings provide a unified microscopic interpretation of ion migration in \ce{MAPbI_{3}} and help reconcile the wide range of activation energies reported experimentally. We show that low activation energies reflect intrinsic bulk diffusion dominated by iodide defects and MA interstitials, whereas higher activation energies reported experimentally likely reflect additional contributions, such as inter-grain transport or formation of complex defects. Overall, our results revise the conventional picture of ionic transport in hybrid perovskites and highlight the importance of collective molecular motion in enabling fast ion migration.

\begin{acknowledgement}

V.T. and S.T. acknowledge funding from Vidi (project no. VI.Vid.213.091) from the Dutch Research Council (NWO).

\end{acknowledgement}

\section{Data Availability}
The training sets for the models used in this article are available in ref.\cite{tyagi_2026_20052556}.

\bibliography{main}

\end{document}


\clearpage

\tableofcontents

\clearpage


\section{Model training} \label{sec:training}
Training structures are sampled from three constant temperature MD runs using the on-the-fly learning procedure as implemented in VASP \cite{Kresse1996efficient,Jinnouchi2019onthefly,Jinnouchi2019phase,bartok2010gaussian,bartok2013onrepresenting}, for both neutral and positively charged methylammonium interstitial (\maintneut and \maintpos), neutral and negatively charged methylammonium vacancy (\mavacneut and \mavacneg), and negatively charged iodide interstitial (\iintneg) and positively charged iodide vacancy (\ivacpos). These runs are performed for \SI{105}{ps} with an MD timestep of \SI{3}{fs} and the mass of \ce{H} increased to \SI{4}{amu}, using \supercell{2}{2}{2} cubic supercells of \ce{MAPbI_{3}} (96 atoms) with one \ce{MA} or \ce{I} point defect in them. Three such runs are performed for each system at temperatures between \SI{500}{K} and \SI{750}{K} (Table~\ref{tab:training_temperatures}) in \textit{NpT} ensembles at $\mathrm{10^{5}}$ Pa pressure. To maintain constant temperature and pressure, Parinello-Rahman dynamics \cite{parrinello1980crystal,parrinello1981polymorphic} is used with friction coefficients set to \SI{3}{ps^{-1}} for all atomic species and lattice degrees of freedom.

\begin{table}[]
    \centering
    \begin{tabular}{|c c c c c|}
        \hline
        System & \textit{n} & Step 1 & Step 2 & Step 3 \\
        \hline
        \maintneut & 1485 & \SI{600}{K} & \SI{700}{K} & \SI{500}{K} \\
        \maintpos & 1596 & \SI{700}{K} & \SI{700}{K} & \SI{500}{K} \\
        \mavacneut & 1339 & \SI{600}{K} & \SI{700}{K} & \SI{500}{K} \\
        \mavacneg & 1328 & \SI{600}{K} & \SI{700}{K} & \SI{500}{K} \\
        \iintneg & 1818 & \SI{600}{K} & \SI{750}{K} & \SI{750}{K} \\
        \ivacpos & 1501 & \SI{600}{K} & \SI{750}{K} & \SI{750}{K} \\
        \hline
    \end{tabular}
    \caption{\textmd{The number of structures (\textit{n}) in the training set and temperatures at which the training runs are performed for different defect systems.}}
    \label{tab:training_temperatures}
\end{table}

The total number of structures sampled for all systems is given in Table~\ref{tab:training_temperatures}. DFT single-point calculations are performed on all these structures in VASP. The projector-augmented wave (PAW) \cite{kresse1999ultrasoft} technique is used to model electron-ion interactions, with the outermost electrons of \ce{I} ($\mathrm{5s^{2}5p^{5}}$), \ce{Pb} ($\mathrm{6s^{2}6p^{2}}$), \ce{H} ($\mathrm{1s^{1}}$), \ce{C} ($\mathrm{2s^{2}2p^{2}}$), and \ce{N} ($\mathrm{2s^{2}2p^{3}}$) treated as valence electrons, and applying the standard VASP PAW potentials. The electronic interactions are modeled using the SCAN+rVV10 functional \cite{Sun2015strongly,Sabatini2013nonlocal}, as it has been shown to reliably describe the dynamics of the \ce{MA} ion and defect formation energies \cite{lahnsteiner2018finite,bokdam2017assesing,xue2021first}. An energy convergence criterion of $\mathrm{10^{-6}}$ \SI{}{eV} is used, along with a \supercell{2}{2}{2} Monkhorst-Pack \textit{k}-point grid\cite{monkhorst1976special} and a kinetic energy cutoff of \SI{500}{eV}.

Three different Allegro models\cite{Batzner2022equivariant,Musaelian2023learning} are trained, one for all charged defects (\maintpos,\mavacneg,\iintneg, and \ivacpos), one for \maintneut, and one for \mavacneut. The defects have these charge states when the Fermi level is well inside the band gap \cite{xue2021first}, and a single force field should be able to capture this. The neutral \maintneut (\mavacneut) can only be found when the Fermi energy is very close or in the conduction (valence) band, and separate force fields have to be constructed to describe these situations.   

All of these models are trained using a radial cutoff of \SI{6.5}{\AA} with interatomic distances projected onto a radial basis using trainable Bessel functions, and 2 tensor product layers. 32 ordered-pair tensor features are used, expanded using spherical harmonics with a maximum angular quantum number ($\mathrm{L_{max}}$) of 2, while preserving full \textit{O}(3) symmetry. The 2-body latent multilayer perceptron (MLP) had the dimensions [128, 256, 512, 1024] and SiLU nonlinearity, and the latent MLP had the dimensions [1024, 1024, 1024] also with SiLU nonlinearity. To predict the pair energies, the final MLP has dimensions [128] without nonlinearity.

For all three models, the training and the validation sets are divided into 80\% and 20\% of the total number of structures (Table~\ref{tab:epochs_and_errors}), respectively, which are shuffled at each epoch during training. The training is performed using the total energy of the system, forces acting on the atoms, and stress tensors. The per-atom MSE loss function is used, with the weight set to 1 for both energy and forces. The Adam optimizer in PyTorch is used with the default parameters $\beta_{1}=0.9$, $\beta_{2}=0.99$, and $\epsilon=10^{-8}$. The learning rate of 0.001 and the batch size of 5 are used. The number of epochs the training of each model ran for is given in Table~\ref{tab:epochs_and_errors}.

\begin{figure}
    \centering
    \includegraphics[]{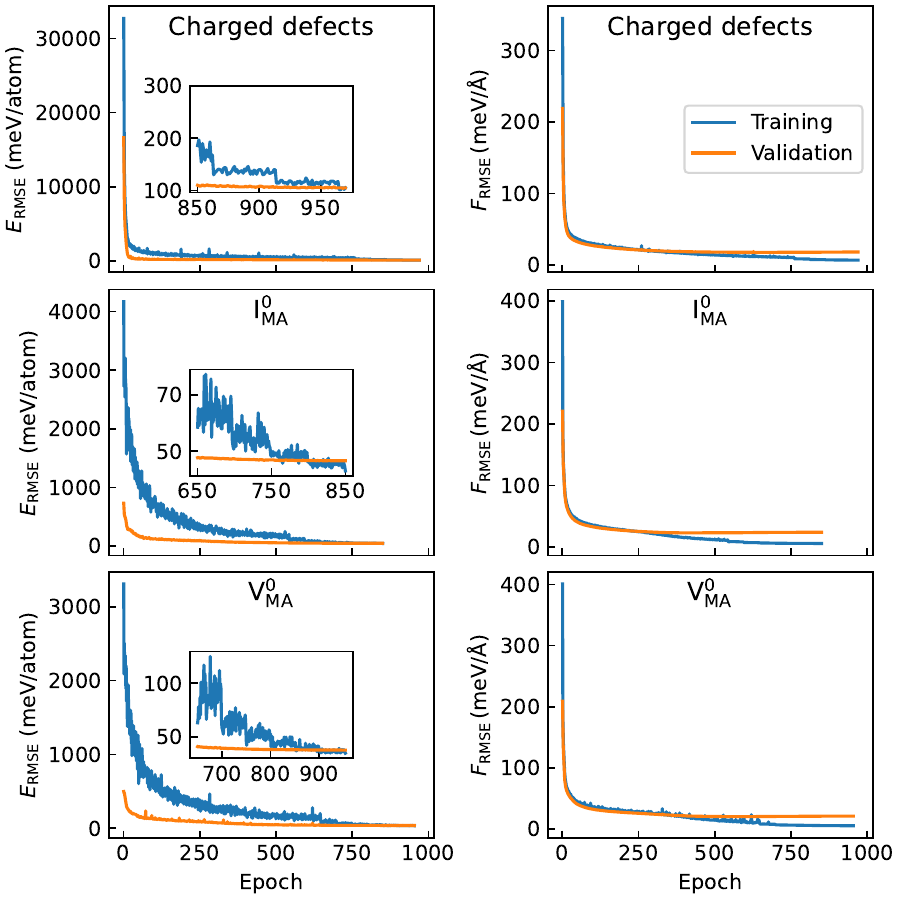}
    \caption{Root mean squared errors in energy ($E_{\mathrm{RMSE}}$) and forces ($F_{\mathrm{RMSE}})$ during model training and model validation of all models.}
    \label{fig:training_errors}
\end{figure}

The root mean squared error (RMSE) in energy and forces during training and validation is given in Figure~\ref{fig:training_errors}. The total number of epochs, and training and validation errors in energy and forces for the final model are given in Table~\ref{tab:epochs_and_errors}.

\begin{table}[]
    \centering
    \begin{tabular}{|c c c c c c c|}
        \hline
        Model & \textit{n} & epochs & \multicolumn{2}{c}{$F_{\mathrm{RMSE}}$(\si{meV/\AA})} & \multicolumn{2}{c|}{$E_{\mathrm{RMSE}}$(\si{meV/atom})} \\
        \multicolumn{3}{|c}{} & Training & Validation & Training & Validation \\
        \hline
        charged defects & 6243 & 970 & 6.5 & 17.9 & 105 & 105 \\
        \maintneut & 1485 & 850 & 5.8 & 24.1 & 43 & 46.6 \\
        \mavacneut & 1339 & 954 & 5.2 & 20.9 & 34.9 & 37.8 \\
        \hline
    \end{tabular}
    \caption{\textmd{The total number of structures in the training and validation sets, number of epochs, training and validation errors in energy and forces for all models.}}
    \label{tab:epochs_and_errors}
\end{table}

\clearpage

\section{Model validation}\label{sec:validation}
\begin{figure}
    \centering
    \includegraphics[]{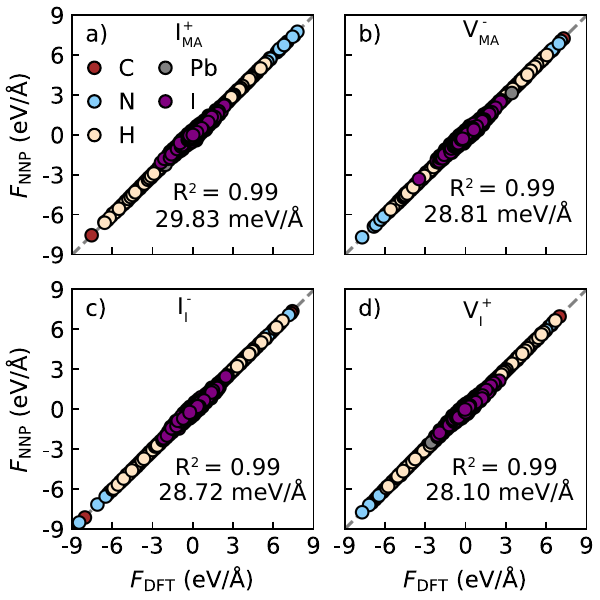}
    \caption{Comparison between forces acting on all atoms calculated by the charged defects NNP with forces calculated using DFT for \supercell{6}{6}{6} cubic supercell of \ce{MAPbI_{3}} with one \ce{MA} interstitial (a), one \ce{MA} vacancy (b), one \ce{I} interstitial (c), or one \ce{I} vacancy, along with corresponding $\mathrm{R^{2}}$ values and mean absolute errors (MAE).}
    \label{fig:validation_forces_charged_all}
\end{figure}

\begin{figure}
    \centering
    \includegraphics[]{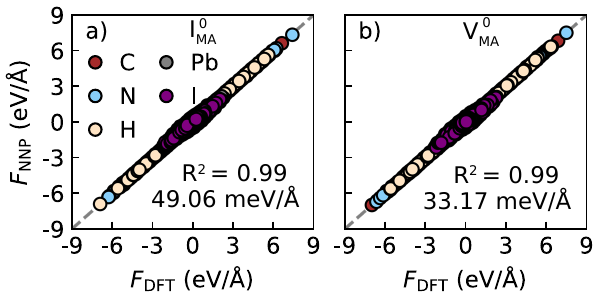}
    \caption{Comparison between forces acting on all atoms calculated by the neutral \ce{MA} interstitial NNP (a) and neutral \ce{MA} vacancy NNP (b) with forces calculated using DFT for \supercell{6}{6}{6} cubic supercell of \ce{MAPbI_{3}} with one \ce{MA} point defect, along with corresponding $\mathrm{R^{2}}$ values and mean absolute errors (MAE).}
    \label{fig:validation_forces_neutral_all}
\end{figure}

To check the accuracy of the model in predicting forces, at least 10 structures are sampled from \SI{0.5}{ns} MD runs at \SI{600}{K} performed using \supercell{6}{6}{6} cubic supercells of \ce{MAPbI_{3}} (2592 atoms) with one \ce{MA} or \ce{I} point defect, and forces calculated using DFT are compared with forces calculated using the NNP for these structures. The forces categorized according to the atomic species acting on all atoms, with the corresponding $\mathrm{R^{2}}$ values and mean absolute errors (MAE) for the charged defects are given in Figure~\ref{fig:validation_forces_charged_all}, and for neutral defects are given in Figure~\ref{fig:validation_forces_neutral_all}. From these comparisons, we note that the $\mathrm{R^{2}}\,=\,0.99$ and $\mathrm{MAE\,\leq\,\SI{29.83}{meV/\angstrom}}$ for all systems. Considering that the forces acting on these atoms are of the order of \SI{1}{eV/\angstrom}, the NNPs are highly accurate in calculating forces acting on all atoms.

\begin{figure}
    \centering
    \includegraphics[]{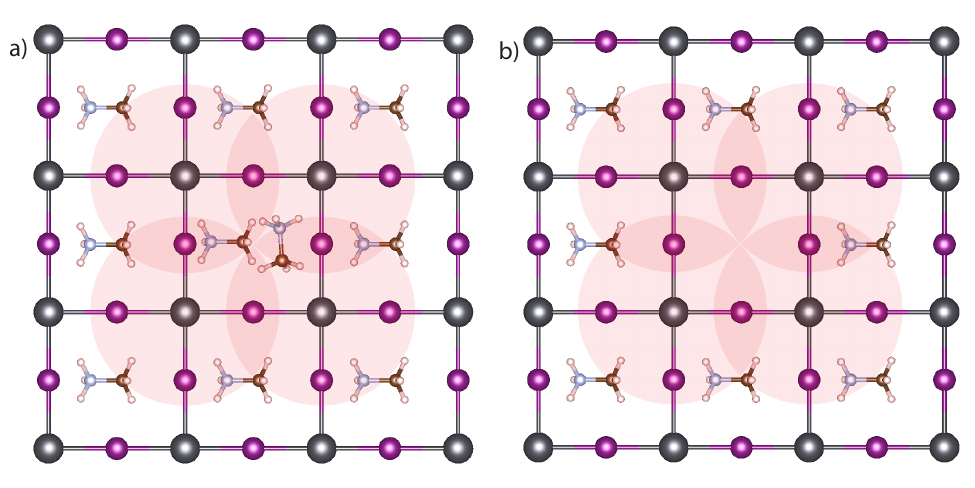}
    \caption{Illustration of how atoms close to the \ce{MA} interstitial (a) and \ce{MA} vacancy (b) are identified for model validation. Here, the red circles indicate the area used for selecting defect environment atoms.}
    \label{fig:MA_defect_identification}
\end{figure}

\begin{figure}
    \centering
    \includegraphics[]{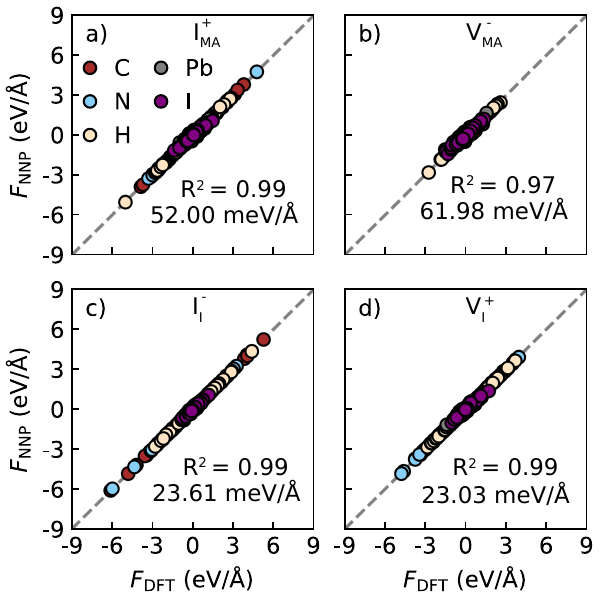}
    \caption{Comparison between forces acting on atoms in defect environment calculated by the charged defects NNP with forces calculated using DFT for a \supercell{6}{6}{6} cubic supercell of \ce{MAPbI_{3}} with one \ce{MA} interstitial (a), one \ce{MA} vacancy (b), one \ce{I} interstitial (c), or one \ce{I} vacancy, along with corresponding $\mathrm{R^{2}}$ values and mean absolute errors (MAE).}
    \label{fig:validation_forces_charged_defect}
\end{figure}

\begin{figure}
    \centering
    \includegraphics[]{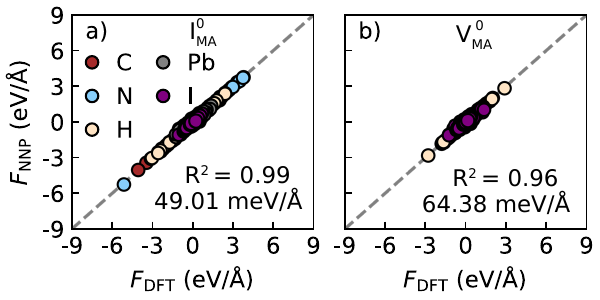}
    \caption{Comparison between forces acting on atoms in defect environment calculated by the neutral \ce{MA} interstitial NNP (a) and neutral \ce{MA} vacancy NNP (b) with forces calculated using DFT for \supercell{6}{6}{6} cubic supercell of \ce{MAPbI_{3}} with one \ce{MA} point defect, along with corresponding $\mathrm{R^{2}}$ values and mean absolute errors (MAE).}
    \label{fig:validation_forces_neutral_defect}
\end{figure}

Forces acting on the atoms close to the defect are also compared. \ce{MA} defect environments are identified by counting the \ce{MA} neighbors of \ce{Pb}, as illustrated in Figure~\ref{fig:MA_defect_identification}.  For the \ce{MA} interstitial (Figure~\ref{fig:MA_defect_identification}a) the environment consists of \ce{Pb} atoms with 9 \ce{MA} neighbors, and for a \ce{MA} vacancy (Figure~\ref{fig:MA_defect_identification}b) the environment consists of \ce{Pb} atoms with 7 \ce{MA} neighbors. The \ce{I} point defect environments are identified using the same procedure described in ref.\cite{tyagi_tracing_2025}. These comparisons are given in Figure~\ref{fig:validation_forces_charged_defect} for charged defects, and in Figure~\ref{fig:validation_forces_neutral_defect} for neutral defects. From these comparisons, we conclude that the NNPs also calculate forces acting on atoms in the defect environment with high accuracy, with the $\mathrm{R^{2}}\,\geq\,0.96$ and $\mathrm{MAE\,\leq\,\SI{64.38}{meV/\angstrom}}$ for all systems.

Along with the forces, the energies calculated with the NNPs are also compared with those calculated using DFT for paths constructed using the climbing image nudged elastic band (CI-NEB) method.\cite{henkelman2000climbing} These paths are constructed using \supercell{2}{2}{1} supercell of \ce{MAPbI_{3}} (16 units) with one \ce{MA} or \ce{I} point defect. Three intermediate images, connected by springs with spring constants $\mathrm{5\,eV/\text{\AA}^{2}}$, are optimized using the same DFT parameters as in Section~\ref{sec:training}, and a \supercell{2}{2}{3} Monkhorst-Pack \textit{k}-grid.

\clearpage

\section{\ce{MAPbI_{3}} phase transition}\label{sec:phase_transition}
To test the method further and to extract the temperature dependence of the volume, a separate NNP is constructed to describe pristine \ce{MAPbI_{3}}. Training structures are sampled from four constant temperature MD runs using the on-the-fly learning procedure as implemented in VASP. These runs are performed for \SI{105}{ps} with an MD timestep of \SI{3}{fs} and the mass of \ce{H} increased to \SI{4}{amu} using \supercell{2}{2}{2} pseudo-cubic supercells of \ce{MAPbI_{3}} in all three perovskite phases. The first run is performed in the cubic phase at \SI{450}{K}, followed by two runs in the tetragonal phase at \SI{250}{K} and \SI{150}{K}, and finally one run in the orthorhombic phase at \SI{150}{K}. All these runs are performed in an \textit{NpT} ensemble with Parinello-Rahman dynamics with friction coefficients set to 3 $\mathrm{ps^{-1}}$ for all atomic species and lattice degrees of freedom.

A total of 1132 structures are sampled, and DFT single-point calculations are performed on all these structures using the same DFT parameters as in SI note ~\ref{sec:training} to generate the final training and validation sets.

\begin{figure}
    \centering
    \includegraphics[]{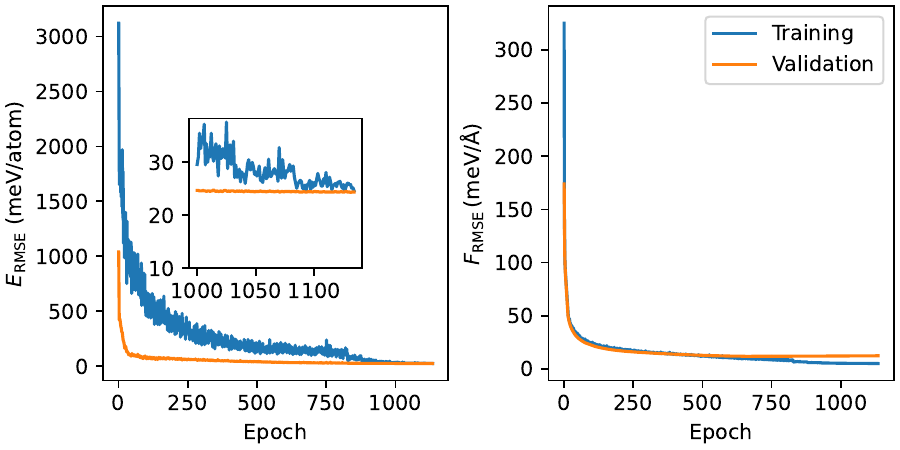}
    \caption{Root mean squared errors in energy ($E_{\mathrm{RMSE}}$) and forces ($F_{\mathrm{RMSE}})$ during model training and model validation of the \ce{MAPbI_{3}} phase transition model.}
    \label{fig:PT_training_errors}
\end{figure}

An Allegro model is trained using the same model parameters as in SI note ~\ref{sec:training}, with the training and validation sets consisting of 905 and 227 structures, respectively, and the training ran for 1134 epochs. The RMSE in energy and forces during model training and model validation are given in Figure ~\ref{fig:PT_training_errors}. The final model has the training errors of \SI{24.5}{meV/atom} in energy and \SI{5.04}{meV/\angstrom} in forces, and validation errors of \SI{24.23}{meV/atom} in energy and \SI{11.7}{meV/\angstrom} in forces.

\begin{figure}
    \centering
    \includegraphics[]{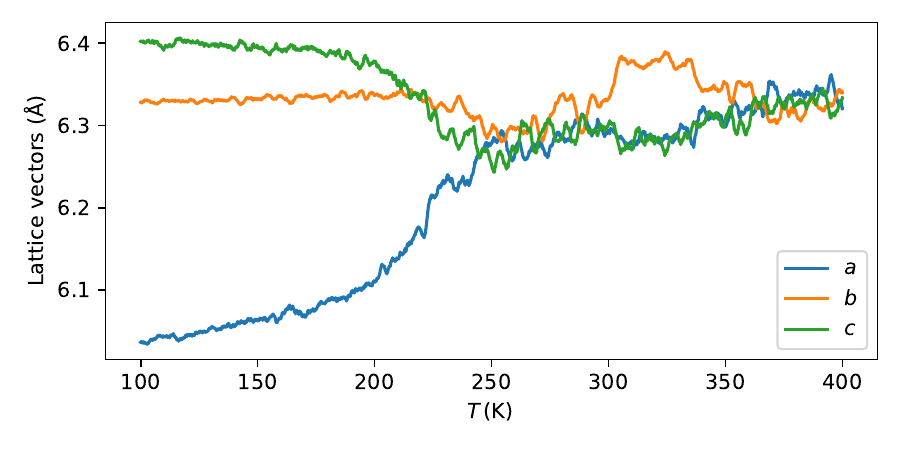}
    \caption{Pseudo-cubic unit cell lattice vectors of \ce{MAPbI_{3}} as functions of temperature in the heating MD runs performed using the phase transition NNP.}
    \label{fig:MAPbI3_heating}
\end{figure}

The accuracy and transferability of this model are validated by performing heating runs from \SI{100}{K} to \SI{400}{K} for \SI{0.6}{ns} using a time step of \SI{1}{fs} on \supercell{3}{3}{2} orthorhombic supercells (72 units) of \ce{MAPbI_{3}}. The window-averaged pseudo-cubic lattice vectors with the window lengths \SI{5}{ps} from this run are given in Figure ~\ref{fig:MAPbI3_heating}. The overall phase diagram closely resembles that observed in experiments \cite{Whitfield2016structures}. 

\begin{figure}
    \centering
    \includegraphics[]{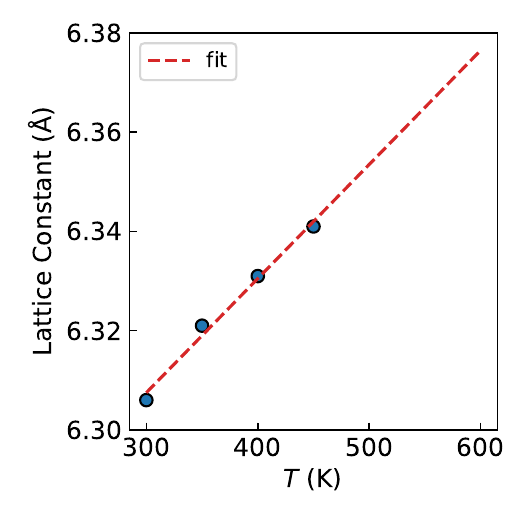}
    \caption{The lattice constant of cubic \ce{MAPbI_{3}} as a function of temperature extracted using the phase transition NNP. The blue points represent calculated values, and the dotted red line represents the fit.}
    \label{fig:MAPbI3_lattice_constant_fit}
\end{figure}

This NNP is used to perform constant temperature MD runs in an \textit{NpT} ensemble using \supercell{6}{6}{6} cubic supercells (216 units) of \ce{MAPbI_{3}} at different temperatures between \SI{300}{K} and \SI{450}{K} with steps of \SI{50}{K} to obtain equilibrated lattice constants. These values are fitted using linear least-squares regression (Figure~\ref{fig:MAPbI3_lattice_constant_fit}) and extrapolated to obtain equilibrated volumes between \SI{500}{K} and \SI{600}{K}.

\clearpage

\section{Production runs and diffusion coefficients}\label{sec:production}
\SI{2}{ns} long MD runs are performed at different temperatures between \SI{500}{K} and \SI{600}{K} using \supercell{6}{6}{6} supercells (216 units of \ce{MAPbI_{3}}) with one \ce{MA} or \ce{I} point defect in LAMMPS\cite{thompson2022lammps}. The lattice constants are obtained using the procedure described in SI note~\ref{sec:phase_transition}, and are given in Table~\ref{tab:MAPbI3_lattice_constants}.

\begin{table}[]
    \centering
    \begin{tabular}{|c c|}
        \hline
        Temperature (K) & Lattice constants ($\text{\AA}$) \\
        \hline
        500 & 6.354 \\
        525 & 6.359 \\
        550 & 6.365 \\
        575 & 6.371 \\
        600 & 6.377 \\
        \hline
    \end{tabular}
    \caption{\textmd{Lattice constants of \ce{MAPbI_{3}} unit cell at different temperatures in cubic phase.}}
    \label{tab:MAPbI3_lattice_constants}
\end{table}

\begin{figure}
    \centering
    \includegraphics[]{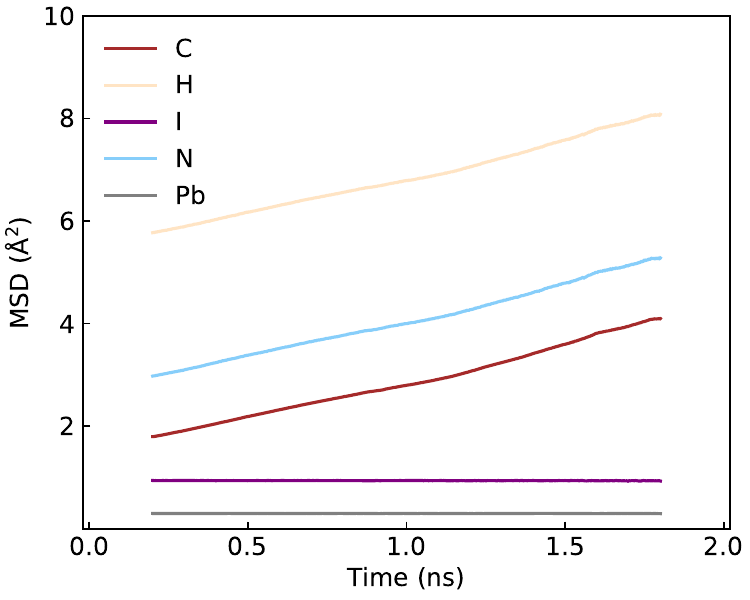}
    \caption{Mean squared displacement (MSD) curves decomposed to the atomic species over simulation time for \maintpos in \ce{MAPbI_{3}} at \SI{550}{K}.}
    \label{fig:msd_ma}
\end{figure}

\begin{figure}
    \centering
    \includegraphics[]{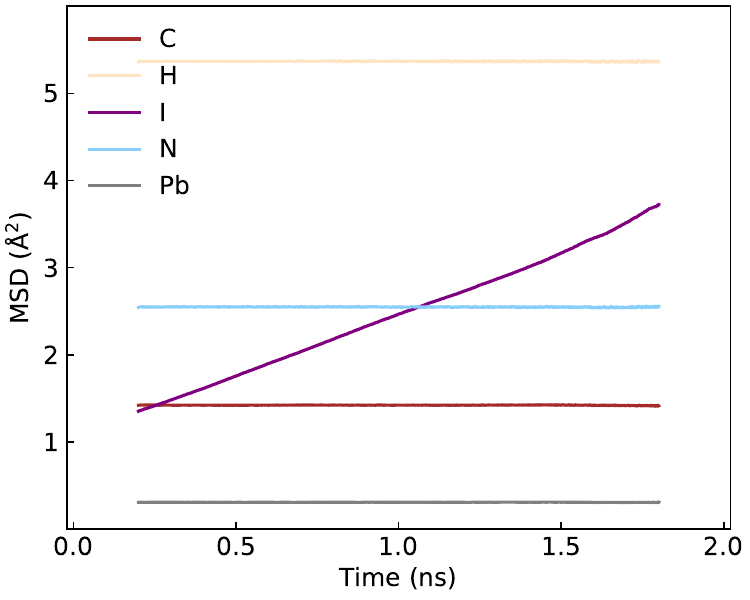}
    \caption{Mean squared displacement (MSD) curves decomposed to the atomic species over simulation time for $\mathrm{I_{I}^{-}}$ in \ce{MAPbI_{3}} at \SI{550}{K}.}
    \label{fig:msd_I}
\end{figure}

The structures are first equilibrated to the target temperature for \SI{100}{ps} with timesteps of \SI{1}{fs} in an \textit{NVT} ensemble using a Langevin thermostat with the relaxation time set to \SI{0.33}{ps}. Following equilibration, \SI{2}{ns} long production runs with timesteps of \SI{1}{fs} are performed in an \textit{NVT} ensemble using a Nose-Hoover thermostat with the relaxation time set to \SI{0.1}{ps}.  To ensure proper sampling of diffusion coefficients, 5 runs are performed at each temperature. To quantify the migration behavior of the defects, the mean squared displacement (MSD) is plotted over time for each atomic species using the MDAnalysis Python library\cite{michaud2011mdanalysis,Maginn2018best}. The MSD plots for \maintpos and $\mathrm{I_{I}^{-}}$ are given in Figure~\ref{fig:msd_ma} and Figure~\ref{fig:msd_I}, respectively. The MSD is calculated using
\begin{equation}
    \text{MSD}(r_{d}) = \Bigl \langle \frac{1}{N}\sum_{i=1}^{N}|r_{d}-r_{d}(t_{0})|^{2}\Bigr \rangle_{t_{0}},
    \label{eqn:equation_MSD}
\end{equation}
where $N$ is the number of atoms of a particular atomic species, and $r_{d}$ are their coordinates in $d$ dimensions (3 for our systems). The Diffusion coefficient $D$ is calculated using
\begin{equation}
    D = \frac{N}{2d}\lim_{t\rightarrow\infty}\frac{d}{dt}\text{MSD}(r_{d}),
    \label{eqn:equation_DC}
\end{equation}
which is proportional to the slope of the MSD curve, where the factor $N$ ensures that the diffusion coefficient is defect concentration independent. The diffusion coefficient of the \ce{MA} molecule is calculated as the average of the diffusion coefficients of \ce{C},\ce{N}, and \ce{H} (Figure~\ref{fig:msd_ma}).

\clearpage

\section{Charge density analysis}\label{sec:charge_density}
\begin{figure}
    \centering
    \includegraphics[]{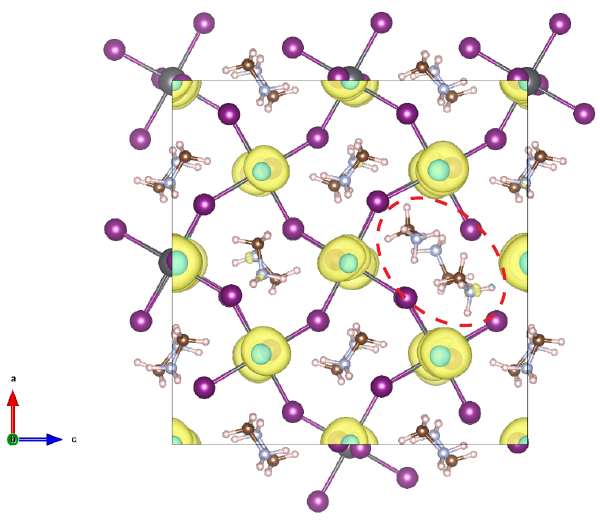}
    \caption{The charge density difference between the $\mathrm{I_{MA}^{0}}$ and $\mathrm{I_{MA}^{+}}$, in the optimized geometry of $\mathrm{I_{MA}^{0}}$. The regions with a negative charge are colored in yellow, and the regions with a positive charge are colored in blue. The defect environment is circled in dotted red lines.}
    \label{fig:rho_difference}
\end{figure}

To check the charge distribution of the extra electron in $\mathrm{I_{MA}^{0}}$, its charge density is compared with that of its positive counterpart. The charge densities of the neutral and positive charge states are calculated using the optimized geometry of $\mathrm{I_{MA}^{0}}$. The difference between the charge densities is visualized in VESTA \cite{momma2008vesta}, and is given in Figure~\ref{fig:rho_difference}. As evident from the figure, the extra electron delocalizes over the whole lattice.

\clearpage
\bibliography{SI}